\documentclass[twocolumn]{cifa}
\usepackage[latin1]{inputenc}
\usepackage[francais]{babel}
\usepackage{graphicx}
\usepackage{amsmath}
\usepackage{amssymb}
\usepackage{color}
\usepackage{amsfonts}
\usepackage{epsf}
\usepackage{epsfig}
\usepackage{subfigure}

\newtheorem{remarque}{\it Remarque\/}

\begin{document}
\title{{\bf{A-t-on vraiment besoin \\ d'un mod\`{e}le probabiliste \\
en ing\'{e}nierie financi\`{e}re?}} \\ ~ \\ {\it Is a probabilistic modeling
\\ really useful in financial engineerinng$\textit{?}$}
\\ ~ \\}
\author{Michel \textsc{Fliess}\textsuperscript{1},
Cedric \textsc{Join}\textsuperscript{2}$^,$\textsuperscript{3},
Fr\'{e}d\'{e}ric \textsc{Hatt}\textsuperscript{4} \\
\normalsize \vskip1em   
\textsuperscript{1}
LIX (CNRS, UMR 7161), \'{E}cole polytechnique, 91128 Palaiseau, France \\
{\tt Michel.Fliess@polytechnique.edu} \\
\textsuperscript{2}
CRAN (CNRS, UMR 7039), Nancy-Universit\'{e}, BP 239, 54506 Vand{\oe}uvre-l\`{e}s-Nancy, France\\
{\tt Cedric.Join@cran.uhp-nancy.fr}
\\
\textsuperscript{3} \'Equipe NON-A, INRIA Lille -- Nord-Europe \\
\textsuperscript{4}Lucid Capital Management, 2 avenue
Charles de Gaulle, BP 351, 2013 Luxembourg, Luxembourg \\
{\tt hatt@lucid-cap.com} \vspace{0.4cm}\begin{flushright} {\small
\og La langue
fran\c{c}aise, sobre et timide, serait encore la derni\`{e}re des langues \\
si la masse de ses bons \'{e}crivains ne l'e\^{u}t pouss\'{e}e au premier rang
\\ en for\c{c}ant son naturel. \fg \\ ~ \\
Antoine de {\sc Rivarol} \\
\textit{Discours sur l'universalit\'{e} de la langue fran\c{c}aise}}
\end{flushright}}

\maketitle

\begin{abstract}~Une vision nouvelle, sans mod\`{e}les math\'{e}matiques
ni outils probabilistes, des chroniques financi\`{e}res permet non
seulement de d\'{e}gager de fa\c{c}on rigoureuse les notions de tendance et
de volatilit\'{e}, mais aussi de fournir des instruments de calculs
efficaces, d\'{e}j\`{a} test\'{e}s avec plein succ\`{e}s en automatique et en
signal. Elle repose sur un th\'{e}or\`{e}me publi\'{e} en 1995 par P. Cartier et
Y. Perrin. On utilise ces r\'{e}sultats pour esquisser une gestion
dynamique des portefeuilles et de strat\'{e}gies, qui fait fi de tout
calcul d'optimisation globale. On pr\'{e}sente de nombreuses simulations
num\'{e}riques.
 \\ ~ \\
{\it Abstract}--- A new standpoint on financial time series, without
the use of any mathematical model and of probabilistic tools, yields
not only a rigorous approach of trends and volatility, but also
efficient calculations which were already successfully applied in
automatic control and in signal processing. It is based on a theorem
due to P. Cartier and Y. Perrin, which was published in 1995. The
above results are employed for sketching a dynamical portfolio and
strategy management, without any global optimization technique.
Numerous computer simulations are presented. \\ ~ \\
\end{abstract}


\newpage

\section{Introduction}
\'Econom\'{e}trie et finance quantitative placent probabilit\'{e}s et
statistiques au c{\oe}ur des fondements th\'{e}oriques et des calculs
pratiques. Ce r\^{o}le, \'{e}galement pr\'{e}pond\'{e}rant dans beaucoup d'autres
sciences appliqu\'{e}es, ne va pas sans poser de redoutables questions
\'{e}pist\'{e}mologiques. Que l'on songe aux r\'{e}flexions, anciennes, de
Keynes \cite{keynes} et Borel \cite{borel}, et, \`{a} celles, plus
r\'{e}centes, de Hacking \cite{hacking} et Jaynes \cite{jaynes}, et,
d'un point de vue plus concret, de Beauzamy \cite{beauzamy}.

La crise financi\`{e}re pr\'{e}sente a mis au centre du d\'{e}bat les critiques
d\'{e}vastatrices de Mandelbrot \cite{mandelbrot} et de certains de ses
\'{e}pigones, comme Taleb \cite{taleb}. En voici un r\'{e}sum\'{e}, sans doute
trop lapidaire: on veut hisser les probabilit\'{e}s sur un socle plus
solide, de fa\c{c}on \`{a} pallier les carences ayant conduit aux impasses
actuelles. Mentionnons, par exemple, la recherche \cite{mandelbrot}
de lois de probabilit\'{e}s \`{a} \og queues \'{e}paisses\footnote{\textit{Fat
tails}, en anglais.} \fg\hspace{0.05cm} pour prendre en compte les
\'{e}v\`{e}nements extr\^{e}mes, rares, recherche, qui de l'aveu m\^{e}me de
\cite{mandelbrot}, n'a pas eu les retomb\'{e}es esp\'{e}r\'{e}es en raison d'un
calibrage malais\'{e}, pour ne pas dire impossible.

Les auteurs pr\^{o}nent, ici, une vision sans probabilit\'{e}s ni
statistiques. La citation suivante, emprunt\'{e}e \`{a} \cite{fes}, en donne
l'esprit:
\\
\noindent\og \textit{Is it not therefore quite na\"{\i}ve to wish to
exhibit well defined probability laws in quantitative finance, in
economics and management, and in other social and psychological
sciences, where the environmental world is much more involved than
in any physical system$?$ In other words {\bf a mathematical theory
of uncertain sequences of events should not necessarily be confused
with probability theory}. To ask if the uncertainty of a ``complex''
system is of probabilistic nature is an undecidable metaphysical
question which cannot be properly answered via experimental means.
It should therefore be ignored.} \fg

Les deux th\`{e}mes suivants fourniront mati\`{e}re \`{a} illustration:
\begin{enumerate}
\item Les \emph{s\'{e}ries temporelles}, ou {\em chronologiques}, appel\'{e}es ici
\emph{chroniques}, sont l'un des piliers de l'\'{e}conom\'{e}trie. Leur
emploi est g\'{e}n\'{e}ralis\'{e}, y compris en finance. Une th\'{e}orie
math\'{e}matique \'{e}l\'{e}gante existe (voir, par exemple,
\cite{gourieroux,hamilton}) dans le cas stationnaire avec des
mod\`{e}les probabilistes lin\'{e}aires. Les op\'{e}rations pour se ramener au
stationnaire, comme la \og d\'{e}saisonnalisation \fg, l'enl\`{e}vement de
tendances, la \og co-int\'{e}gration \fg, sont d\'{e}licates et ne
conduisent pas \`{a} des pr\'{e}dictions vraiment satisfaisantes. Le recours
\`{a} des mod\`{e}les non lin\'{e}aires n'am\'{e}liore gu\`{e}re la situation.
\item La \og th\'{e}orie moderne du portefeuille\footnote{\textit{Modern portfolio theory},
en anglais, souvent abr\'{e}g\'{e} en \textit{MPT}.} \fg \ de Markowitz
(\cite{marko1,marko2}), qui vise un bon compromis entre rendements
et risques, est l'une des premi\`{e}res, sinon la premi\`{e}re,
manisfestations de l'emprise des probabilit\'{e}s en finance
quantitative\footnote{Voir, par exemple, les excellents cours
\cite{bertrand,bodie}.}. Sa mise en {\oe}uvre concr\`{e}te, qui repose sur
une matrice de variances/covariances, est lourde et cache des
chausse-trapes vicieux\footnote{Voir, par exemple, \cite{lefeuvre}
pour un r\'{e}sum\'{e} court, mais lumineux, de ces difficult\'{e}s, et des
r\'{e}f\'{e}rences compl\'{e}mentaires.}. Le passage du statique \`{a} une gestion
dynamique fait souvent appel \`{a} la commande optimale stochastique,
promue par des c\'{e}l\'{e}brit\'{e}s de la finance math\'{e}matique, telles
Samuelson \cite{samuelson} et Merton \cite{merton}. La complexit\'{e}
formidable de l'optimisation math\'{e}matique, surtout si elle est
dynamique et stochastique, nuit \`{a} toute implantation num\'{e}rique, \`{a}
moins de simplifications drastiques. C'est de plus une gageure, \`{a}
notre avis, que vouloir d\'{e}crire l'\'{e}volution des prix et des
rendements par des \'{e}quations diff\'{e}rentielles stochastiques, qui
sont, rappelons-le, un pilier de la math\'{e}matique financi\`{e}re depuis
quarante ans (voir, par exemple,
\cite{dana,hull,merton,roncalli,wilmott})\footnote{Signalons
d'autres tentatives, \og moins th\'{e}oriques \fg, comme, par exemple,
les r\'{e}seaux de neurones \cite{liu}.}.
\end{enumerate}

C'est la renonciation m\^{e}me \`{a} un mod\`{e}le pr\'{e}cis, qui nous indique la
voie pour surmonter ces difficult\'{e}s. Elle repose sur
\begin{itemize}
\item l'existence de moyennes, ou tendances\footnote{L'emploi du terme anglais \textit{trend}
est courant.}, pour les chroniques financi\`{e}res, qui
\begin{itemize}
\item est une hypoth\`{e}se fondamentale en \emph{analyse
technique}\footnote{L'analyse technique (voir, par exemple,
\cite{bechu,kirk}), souvent m\'{e}pris\'{e}e par les th\'{e}oriciens de la
finance quantitative, est d'un emploi courant chez les praticiens.},
\item a \'{e}t\'{e} d\'{e}montr\'{e}e pour la premi\`{e}re fois, apparemment, en
\cite{fes};
\end{itemize}
\item une approche pr\'{e}cise et exploitable de la \emph{volatilit\'{e}} \cite{douai},
qui manquait.
\end{itemize}
Ces notes, qui esquissent un point de vue nouveau sur la gestion
dynamique sont \`{a} rapprocher de la \og commande sans mod\`{e}le \fg, due
\`{a} deux des auteurs (\cite{modelfree1,marseille}). Les succ\`{e}s
remarquables d\'{e}j\`{a} obtenus (voir les r\'{e}f\'{e}rences de \cite{marseille})
s'expliquent par l'inutilit\'{e} d'un \og bon \fg \ mod\`{e}le math\'{e}matique,
le plus souvent impossible \`{a} \'{e}crire \`{a} cause de la complexit\'{e} des
ph\'{e}nom\`{e}nes physiques, comme le frottement, et des perturbations
externes inconnues, pour obtenir des lois de commande performantes
et faciles \`{a} implanter.

Le {\S} \ref{chro} reprend la vision, n\'{e}e en \cite{fes}, des
chroniques, bas\'{e}e sur le th\'{e}or\`{e}me de Cartier-Perrin \cite{cartier}.
Le {\S} \ref{dyn} esquisse une approche enti\`{e}rement nouvelle de la
gestion dynamique de portefeuilles et de strat\'{e}gies, qui \'{e}vite les
calculs lourds d'optimisation. Des simulations num\'{e}riques illustrent
et valident notre d\'{e}marche. La conclusion du {\S} \ref{conclusion}
discute bri\`{e}vement des filtres de Kalman.

\section{Chroniques\protect\footnote{On
pr\'{e}f\`{e}re, comme d\'{e}j\`{a} dit dans l'introduction, cette terminologie,
plus \'{e}l\'{e}gante, \`{a} celle de \emph{s\'{e}ries temporelles}, ou
\emph{chronologiques}.}}\label{chro}

\subsection{Pr\'{e}liminaires}
\subsubsection{Analyse technique} On sait que tout signal
\og utile \fg \ est noy\'{e} dans du bruit. S'il est additif, cela
correspond \`{a}
\begin{equation}\label{obs}
\text{signal observ\'{e}} = \text{signal utile} + \text{bruit}
\end{equation}
Trouver comment r\'{e}cup\'{e}rer les informations pertinentes, gr\^{a}ce \`{a} des
m\'{e}thodes efficaces de d\'{e}bruitage et d'estimation, est une des t\^{a}ches
essentielles des sciences de l'ing\'{e}nieur et des math\'{e}matiques
appliqu\'{e}es. En finance, seule l'\emph{analyse technique}
(\cite{bechu,kirk}) se rapproche de ce point de vue car elle voit
toute chronique des prix d'un actif comme des fluctuations rapides
autour d'une \emph{tendance}\footnote{On emploie souvent le terme
anglais \textit{trend}.}. Alors, \eqref{obs} devient:
\begin{equation}\label{prix}
\text{prix} = \text{tendance} + \text{fluctuations rapides}
\end{equation}
Les math\'{e}matiques financi\`{e}res actuelles, par contre, qui insistent
sur le fait que les prix suivent une marche al\'{e}atoire\footnote{Ces
marches al\'{e}atoires se rattachent \`{a} l'\og hypoth\`{e}se de l'efficience
du march\'{e} \fg, \emph{efficient market hypothesis} en anglais, due \`{a}
Fama \cite{fama}. Une litt\'{e}rature consid\'{e}rable est consacr\'{e}e \`{a} cette
question capitale. On en trouve un r\'{e}sum\'{e} dans certains cours cit\'{e}s
en bibliographie. Voir, par exemple, \cite{bodie,wilmott}.}, nient
ces tendances. L'analyse technique, fort appr\'{e}ci\'{e}e de maints
g\'{e}rants\footnote{On aurait pu user du terme anglais \textit{trader},
aujourd'hui universel.}, est donc rejet\'{e}e par la finance th\'{e}orique
\og moderne \fg.

\subsubsection{Vers une nouvelle th\'{e}orie des chroniques}
La th\'{e}orie des chroniques, telle qu'on
\begin{itemize}
\item la trouve aujourd'hui dans des cours comme
\cite{gourieroux,hamilton},
\item l'utilise non seulement en finance
quantitative, mais aussi en \'{e}conom\'{e}trie et dans bien d'autres
domaines des sciences appliqu\'{e}es,
\end{itemize}
ignore ces tendances\footnote{Le mot \textit{tendance} poss\`{e}de en
\cite{gourieroux,hamilton}, et dans tous les cours actuels sur les
chroniques, un autre sens.}. Or ces tendances existent d'apr\`{e}s le
th\'{e}or\`{e}me de Cartier-Perrin \cite{cartier}, publi\'{e} en 1995, qui est
au c{\oe}ur de la refondation de l'analyse des chroniques, entam\'{e}e
depuis \cite{fes}. On a pu ainsi r\'{e}examiner
(\cite{malo2,beta,delta,douai}) bien des points d'ing\'{e}nierie
financi\`{e}re: nouveaux indicateurs, coefficient b\^{e}ta de risque,
volatilit\'{e}, couverture, etc. Ce th\'{e}or\`{e}me doit, \`{a} notre avis, \^{e}tre
compris comme un r\'{e}sultat important et nouveau de la th\'{e}orie des
fonctions d'une variable r\'{e}elle \cite{bourbaki}. Il est, comme
rappel\'{e} plus bas, exprim\'{e} dans le langage de l'\emph{analyse non
standard}, trop ignor\'{e}. Insistons, ici, sur l'absence de toute loi
de probabilit\'{e}s pour \'{e}tablir la d\'{e}composition \eqref{prix}.

\begin{remarque}Notons que des travaux isol\'{e}s sur les chroniques,
comme le livre d'Andersen \cite{andersen}, vieux de plus de $80$ ans
et donc ant\'{e}rieur \`{a} la doxa dominante, sont moins \'{e}trangers aux
tendances. Le r\'{e}cent et excellent manuel de M\'{e}lard \cite{melard}
tranche aussi avec la plupart des cours universitaires disponibles.
\end{remarque}
\subsubsection{Aspects calculatoires}
Les adeptes de l'analyse techniques savent depuis longtemps (voir,
par exemple, \cite{bechu,kirk}) qu'une bonne fa\c{c}on de d\'{e}gager la
tendance \`{a} partir de \eqref{prix} gagne \`{a} s'inspirer de la pratique
des ing\'{e}nieurs pour traiter \eqref{obs}. Nos m\'{e}thodes de d\'{e}bruitage
et de d\'{e}rivation num\'{e}rique (voir \cite{nl,mboup}), de nature
alg\'{e}brique et test\'{e}es avec plein succ\`{e}s dans de multiples exemples
concrets (voir, par exemple, \cite{Vil1,Vil2}), am\'{e}liorent les
\emph{moyennes mobiles}, courantes en analyse technique.

\subsection{Analyse non standard et th\'{e}or\`{e}me de Cartier-Perrin}\label{nsa}
\subsubsection{G\'{e}n\'{e}ralit\'{e}s} L'\emph{analyse non standard},
invent\'{e}e par Robinson \cite{robinson} il y a cinquante ans,
accomplit un r\^{e}ve ancien en donnant une base enfin rigoureuse, gr\^{a}ce
\`{a} la logique math\'{e}matique, aux notions d'\og infiniment petit \fg \
et d'\og infiniment grand \fg. On en doit \`{a} Nelson \cite{bams} une
pr\'{e}sentation plus claire et plus accessible, explicit\'{e}e en
\cite{diener1,diener2,robert}.
\subsubsection{D\'{e}finition des chroniques}
Soit l'intervalle $[0, 1] \subset \mathbb{R}$. Introduisons, comme
souvent en analyse non standard, la \emph{discr\'{e}tisation
infinit\'{e}simale}
$${\mathfrak{T}} = \{ 0 = t_0 < t_1 < \dots < t_\nu = 1 \}$$
o\`{u} $t_{i+1} - t_{i}$, $0 \leq i < \nu$, est {\em infinit\'{e}simal},
c'est-\`{a}-dire \og tr\`{e}s petit \fg. Une \emph{chronique} $X(t)$ est une
fonction $X: {\mathfrak{T}} \rightarrow \mathbb{R}$.

\subsubsection{Int\'{e}grabilit\'{e} et continuit\'{e}}
La {\em mesure de Lebesgue} sur $\mathfrak{T}$ est la fonction $m$
d\'{e}finie sur $\mathfrak{T} \backslash \{1\}$ par $m(t_{i}) = t_{i+1} -
t_{i}$. La mesure d'un intervalle $[c, d[ \subset \mathfrak{I}$, $c
\leq d$, est sa longueur $d -c$.  Posons
$$\int_{[c, d[} fdm = \sum_{t \in [c, d[} f(t)m(t)$$
pour l'int\'{e}grale sur $[c, d[$ de la fonction $f: \mathfrak{I}
\rightarrow \mathbb{R}$. La fonction $f: \mathfrak{T} \rightarrow
\mathbb{R}$ est dite $S$-{\em int\'{e}grable} si, et seulement si, pour
tout intervalle $[c, d[$, l'int\'{e}grale $\int_{[c, d[} |f| dm$ est
\emph{limit\'{e}}, et infinit\'{e}simal, si $d - c$ l'est.

La fonction $f$ est dite $S$-{\em continue} en $t_\iota \in
\mathfrak{T}$ si, et seulement si, $t_\iota \simeq \tau
\Longrightarrow f(t_\iota) \simeq f(\tau)$\footnote{$x \simeq y$
signifie que $x - y$ est infinit\'{e}simal.}. La fonction $f$ est dite
{\em presque continue} si, et seulement si, elle est $S$-continue
sur $\mathfrak{T} \setminus R$, o\`{u} $R$ est {\em
rare}\footnote{L'ensemble $R$ est dit {\em rare} si, et seulement
si, pour tout r\'{e}el standard $\alpha > 0$, il existe un ensemble
interne $B \supset A$ tel que $m(B) \leq \alpha$.}. On dit que $f$
est \emph{Lebesgue-int\'{e}grable} si, et seulement si, elle est
$S$-int\'{e}grable et presque continue.
\subsubsection{Fluctuations rapides}\label{rap}
Une fonction $h: \mathfrak{T} \rightarrow \mathbb{R}$ est dite \`{a}
{\em fluctuations}, ou {\em oscillations}, \emph{rapides} si, et
seulement si, elle est
\begin{itemize}
\item $S$-int\'{e}grable,
\item $\int_A h dm$ est infinit\'{e}simal pour tout $A$ {\em quadrable}\footnote{
$A$ est quadrable \cite{cartier} si sa fronti\`{e}re est rare.}.
\end{itemize}
\subsubsection{D\'{e}composition de Cartier-Perrin} Toute chronique $S$-int\'{e}grable $X: \mathfrak{T}
\rightarrow \mathbb{R}$ v\'{e}rifie la d\'{e}composition de
Cartier-Perrin\footnote{Voir \cite{lobry} pour une pr\'{e}sentation \og
classique \fg, c'est-\`{a}-dire sans analyse non standard.}:
\begin{equation}\label{decomposition}
\boxed{X(t) = E(X)(t) + X_{\tiny{\rm fluctuation}}(t)}
\end{equation}
o\`{u}
\begin{itemize}
\item $E(X)(t)$, qui est Lebesgue-int\'{e}grable, est
l'\emph{esp\'{e}rance}, aussi appel\'{e}e \emph{tendance}, ou encore, en
suivant la terminologie am\'{e}ricaine, \emph{trend};
\item $X_{\tiny{\rm fluctuation}}(t)$ est \`{a} fluctuations rapides.
\end{itemize}
La d\'{e}composition \eqref{decomposition} est unique \`{a} un infiniment
petit additif pr\`{e}s.

\subsubsection{Variances et covariances}
D\'{e}finir, dans ce cadre, l'analogue de la (co)variance, et, donc, de
la \emph{volatilit\'{e}} est imm\'{e}diat (voir \cite{douai} pour plus de
d\'{e}tails):
\begin{enumerate}
\item La \emph{covariance} de deux chroniques $X(t)$ et $Y(t)$ est
\begin{eqnarray*}
\mbox{\rm cov}(XY)(t) & = & E\left((X - E(X))(Y - E(Y)) \right)(t)
\\ & \simeq & E(XY)(t) - E(X)(t) \times E(Y)(t)
\end{eqnarray*}
\item La \emph{variance} de la chronique $X(t)$ est
\begin{eqnarray*}\label{var}
\mbox{\rm var}(X)(t) & = & E\left((X - E(X))^2 \right)(t) \\ &
\simeq & E(X^2)(t) - \left(E(X)(t)\right)^2
\end{eqnarray*}
\item La \emph{volatilit\'{e}} de $X(t)$ est l'\'{e}cart-type correspondant:
\begin{equation}\label{vol}
\mbox{\rm vol}(X)(t) = \sqrt{\mbox{\rm var}(X)(t)}
\end{equation}
\end{enumerate}
G\'{e}n\'{e}raliser aux moments d'ordre sup\'{e}rieur est trivial.

\subsection{D\'{e}bruitage et estimation \label{par_debest}}
Le d\'{e}bruitage, c'est-\`{a}-dire, ici, l'att\'{e}nuation des fluctuations
rapides, d\'{e}coule de leur d\'{e}finition m\^{e}me en {\S} \ref{rap}:
l'int\'{e}gration et, plus g\'{e}n\'{e}ralement, tout filtre passe-bas mettent
les esp\'{e}rances des chroniques en \'{e}vidence.

La d\'{e}termination des d\'{e}riv\'{e}es des esp\'{e}rances revient \`{a} la d\'{e}rivation
num\'{e}rique de signaux bruit\'{e}s. C'est, on le sait, un probl\`{e}me d'une
grande importance, ayant suscit\'{e} une litt\'{e}rature consid\'{e}rable en
math\'{e}matiques appliqu\'{e}es et en ing\'{e}nierie. On r\'{e}sume grossi\`{e}rement,
ici, une approche nouvelle \cite{mboup}, d\'{e}but\'{e}e en \cite{compr},
qui a d\'{e}j\`{a} modifi\'{e} notre compr\'{e}hension des questions d'observation,
d'identification param\'{e}trique et de diagnostic en automatique non
lin\'{e}aire \cite{nl}. La possibilit\'{e} d'utiliser des int\'{e}grales pour
estimer les d\'{e}riv\'{e}es remonte au moins, comme not\'{e}e en \cite{lille},
\`{a} Lanczos \cite{lanczos}:
$$\frac{3}{2h^3} \int_{-h}^{h} \tau x(t +
\tau) d\tau = \dot{x}(t) + O(h^2)$$

Soit, pour illustrer ce qui pr\'{e}c\`{e}de, $x(t)$ un signal dont on veut
estimer la d\'{e}riv\'{e}e premi\`{e}re. Approchons $x(t)$ autour de $t = 0$ par
le polyn\^{o}me de Taylor tronqu\'{e} jusqu'\`{a} l'ordre de la d\'{e}rivation
souhait\'{e}e, ici $1$.

Soit, pour simplifier, un signal polyn\^{o}mial de degr\'{e} $1$:
$$p(t)=a_{0}+a_{1}t$$
Avec les notations classiques du calcul
op\'{e}rationnel ({\it cf.} \cite{yosida}), il vient, pour $t \geq 0$,
\begin{equation*}\label{eq5}
P(s)=\frac{a_{0}}{s}+\frac{a_{1}}{s^{2}}
\end{equation*}
Des calculs \'{e}l\'{e}mentaires m\`{e}nent \`{a}
\begin{equation*}\label{eq7}
P(s)+s\frac{\text{d}P(s)}{\text{d}s}=-\frac{a_{1}}{s^{2}}.
\end{equation*}
Avant de revenir au domaine temporel, une multiplication par
$s^{-N}$, avec $N>1$, $N=2$ par exemple, est n\'{e}cessaire pour \'{e}viter
les d\'{e}rivations par rapport au temps et obtenir uniquement des
int\'{e}grales:
\begin{equation*}\label{eq8}
s^{-2}P(s)+s^{-1}\frac{\text{d}P(s)}{\text{d}s}=-s^{-4}a_{1}.
\breve{}\end{equation*}
On revient au domaine temporel en rappelant
({\it cf.} \cite{yosida}) que $\frac{d}{ds}$ correspond \`{a} la
multiplication par $-t$:
\begin{equation}
\begin{split}
a_{1}&=\frac{6\Bigg(\displaystyle\int_{t_{0}}^{t}\tau
x(\tau)\text{d}\tau-\int_{t_{0}}^{t}\int_{t_{0}}^{\tau}x(\kappa)\text{d}\kappa\text{d}\tau\Bigg)}{\displaystyle
t^{3}}\\&=\frac{6\Bigg(\displaystyle\int_{t_{0}}^{t}\tau
x(\tau)\text{d}\tau-\int_{t_{0}}^{t}(t-\tau)x(\tau)\text{d}\tau\Bigg)}{\displaystyle
t^{3}}\\&=\frac{6\displaystyle\int_{t_{0}}^{t}\big(\tau
x(\tau)-(t-\tau)x(\tau)\big)\text{d}\tau}{\displaystyle t^{3}}
\end{split}\label{eq15}
\end{equation}
Les \'{e}tapes pr\'{e}c\'{e}dentes ne sont pas univoques et une multitude de
formules de type \eqref{eq15} sont possibles. La g\'{e}n\'{e}ralisation \`{a}
des polyn\^{o}mes de degr\'{e} quelconque est imm\'{e}diate. On estime les
d\'{e}riv\'{e}es d'une fonction analytique, en tronquant son d\'{e}veloppement
de Taylor, c'est-\`{a}-dire en se ramenant au cas pr\'{e}c\'{e}dent. La
discr\'{e}tisation conduit \`{a} un filtre num\'{e}rique.
\begin{remarque}\label{remarque1}
Voir \cite{mboup} et, aussi, \cite{lille} pour des d\'{e}veloppements
importants sur les retards associ\'{e}s \`{a} ces estimateurs. Voir, par
exemple, \cite{gretsi} pour une implantation pratique.
\end{remarque}

\subsection{Rendements}
\subsubsection{G\'{e}n\'{e}ralit\'{e}s}
Le \emph{rendement logarithmique}\footnote{Renvoyons \`{a}
\cite{douai} pour plus de d\'{e}tails.}, sur l'intervalle de temps
$\Delta T
> 0$, de l'actif $\mathfrak{A}$, dont le prix \`{a}
l'instant $t$ est $X(t)$,  est la chronique $R_{\Delta T}$ d\'{e}finie
par
\begin{eqnarray*}
R_{\Delta T} ({\mathfrak{A}}) (t) & = & \ln \left( \frac{X(t)}{X(t -
\Delta T)} \right) \\ & = & \ln X(t) - \ln X(t - \Delta T)
\end{eqnarray*}
D\'{e}finissons le rendement logarithmique \emph{normalis\'{e}} par
\begin{equation*}\label{nor}
r_{\Delta T} ({\mathfrak{A}}) (t) = \frac{R(t)}{\Delta T}
\end{equation*}
La \emph{moyenne} de $r_{\Delta T} ({\mathfrak{A}}) (t)$ est
\begin{equation}\label{aver}
\bar{r}_{\Delta T}({\mathfrak{A}})(t) = \frac{E (\ln X) (t) - E (\ln
X) (t - \Delta T)}{\Delta T}
\end{equation}
Si $E (\ln X)$ est d\'{e}rivable,
\begin{equation*}\label{r}
\bar{r} ({\mathfrak{A}}) (t) = \frac{d}{dt} E (\ln X) (t)
\end{equation*}
est le {\em rendement logarithmique instantan\'{e}}.
\subsubsection{Volatilit\'{e} d'un actif}
La \emph{volatilit\'{e} historique}, ou, plus bri\`{e}vement, la
\emph{volatilit\'{e}}, de $\mathfrak{A}$ est
\begin{equation}\label{vola}
{\text{\bf vol}}_{\Delta T}   ({\mathfrak{A}})(t)  = \sqrt{E \left(
r_{\Delta T} ({\mathfrak{A}}) - \bar{r}_{\Delta T} ({\mathfrak{A}})
\right)^2  (t)}
\end{equation}
D'o\`{u}
$$
{\text{\bf vol}}_{\Delta T}  ({\mathfrak{A}})(t) \simeq
\sqrt{E((r_{\Delta T}({\mathfrak{A}}))^2)(t) - (\bar{r}_{\Delta T}
({\mathfrak{A}}) (t))^2}
$$

\subsubsection{Ratio de Sharpe} Le \emph{ratio de Sharpe}
(\cite{sharpe1,sharpe2})
d'un actif ${\mathfrak{A}}$ est une mesure tr\`{e}s populaire de la
performance d'un portefeuille (voir, par exemple,
\cite{bodie,roncalli,wilmott}). Il s'\'{e}crit, ici,
\begin{equation}\label{sharp}
\text{SR}_{\Delta T}({\mathfrak{A}}) (t) = \frac{\bar{r}_{\Delta
T}({\mathfrak{A}}) (t)}{{\text{\bf vol}}_{\Delta T}({\mathfrak{A}})
(t) }
\end{equation}
On souhaite un ratio de Sharpe \'{e}lev\'{e}: fort rendement, c'est-\`{a}-dire
num\'{e}rateur grand, et risque faible, c'est-\`{a}-dire d\'{e}nominateur petit.

\subsection{Illustrations num\'{e}riques\label{ill_num}}
Soit le cours journalier de l'or du 30 septembre 1991 au 27 ao\^{u}t
2010.
\subsubsection{Comparaison avec une moyenne glissante classique}
\begin{figure*}[!ht]
\centering {\subfigure[\footnotesize Cours de l'or (--) et moyenne
glissante classique (--); esp\'{e}rance propos\'{e}e (- -) ]{
\rotatebox{-0}{\resizebox{!}{6.5cm}{%
   \includegraphics{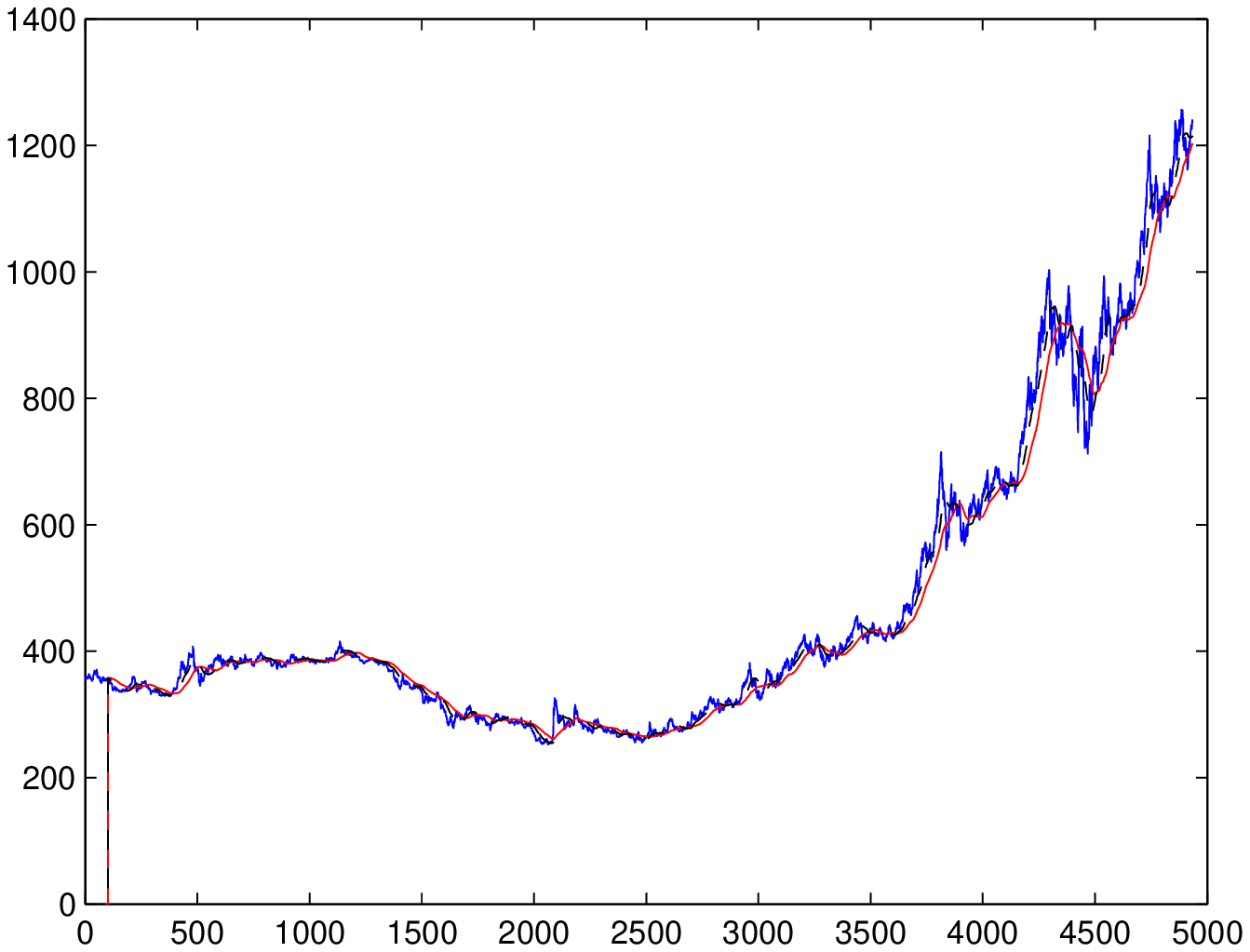}}}}}
{\subfigure[\footnotesize Zoom de \ref{fig_trend}-(a)]{
\rotatebox{-0}{\resizebox{!}{6.5cm}{\includegraphics{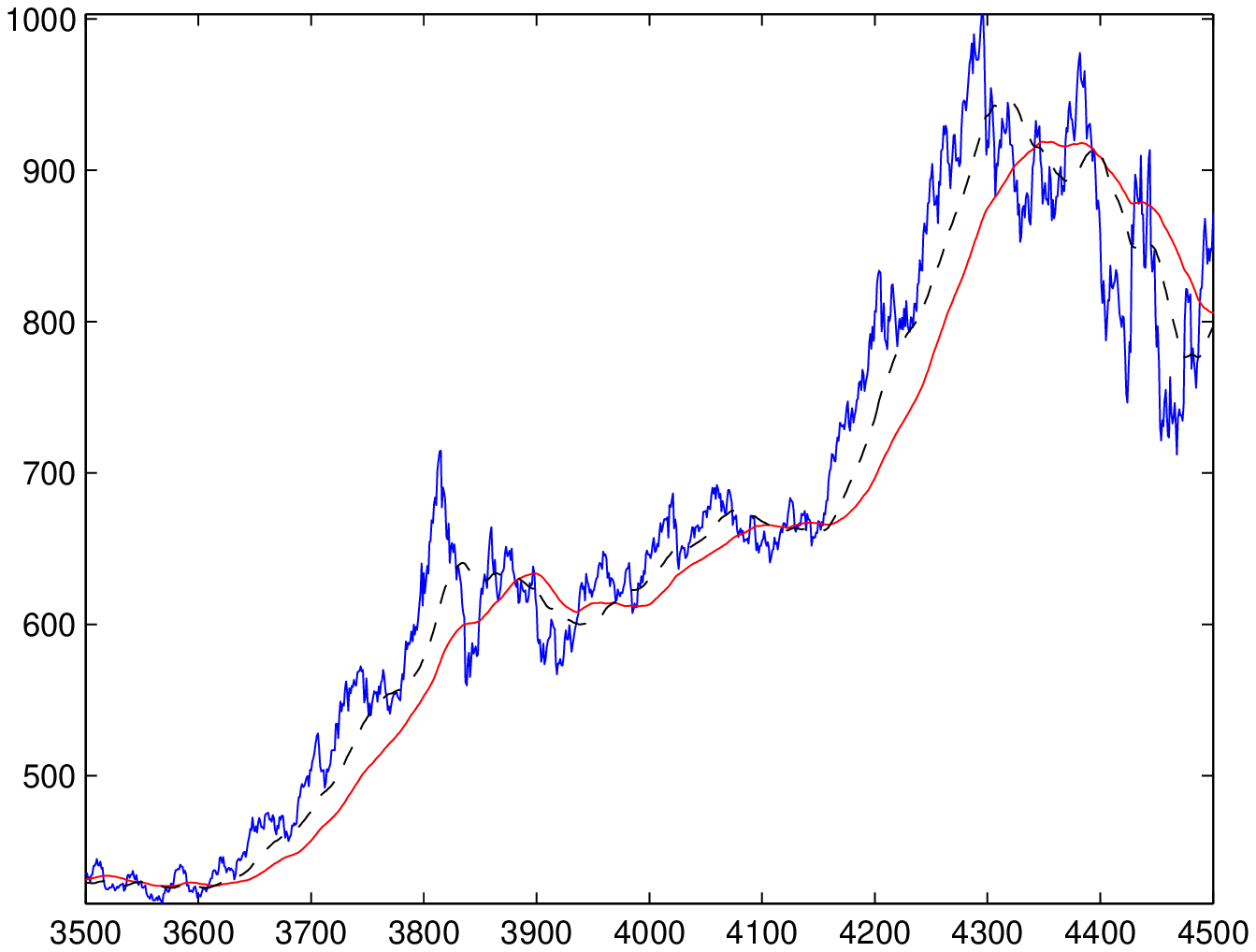}}}}}
 \caption{Calculs d'esp\'{e}rances \label{fig_trend}}
\end{figure*}

La figure \ref{fig_trend} compare une moyenne mobile avec la
m\'{e}thodologie r\'{e}sum\'{e}e au {\S} \ref{par_debest}. Quoique toutes deux
utilisent 100 points, c'est-\`{a}-dire 100 jours, la seconde diminue
\'{e}norm\'{e}ment le retard d'estimation, sans affaiblir le d\'{e}bruitage.

\subsubsection{D\'{e}rivation num\'{e}rique}
\begin{figure*}[!ht]
\centering {\subfigure[\footnotesize D\'{e}rivations num\'{e}riques
classique (--) et nouvelle (- -) ]{
\rotatebox{-0}{\resizebox{!}{6.5cm}{%
   \includegraphics{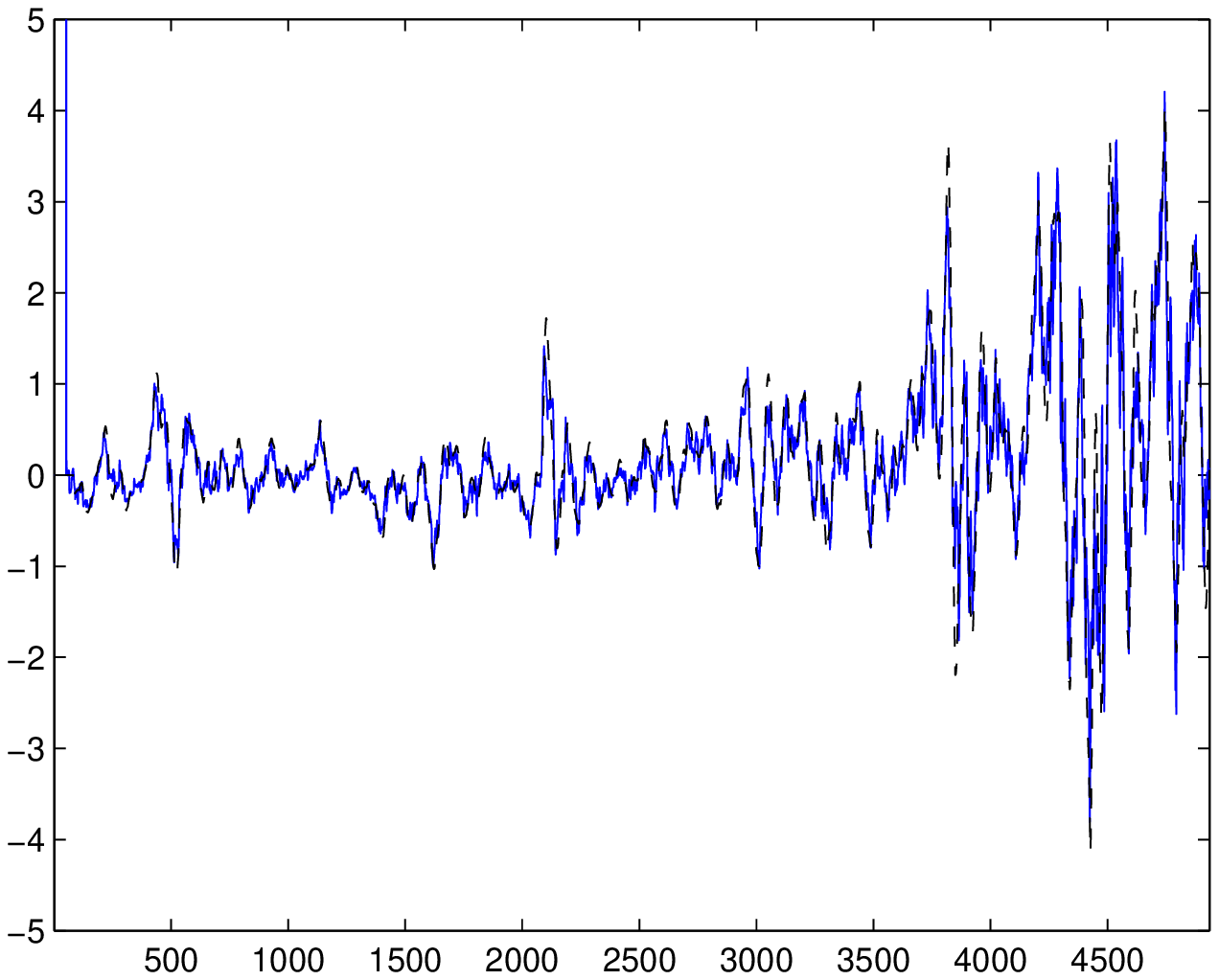}}}}}
{\subfigure[\footnotesize Zoom de \ref{fig_diff}-(a)]{
\rotatebox{-0}{\resizebox{!}{6.5cm}{\includegraphics{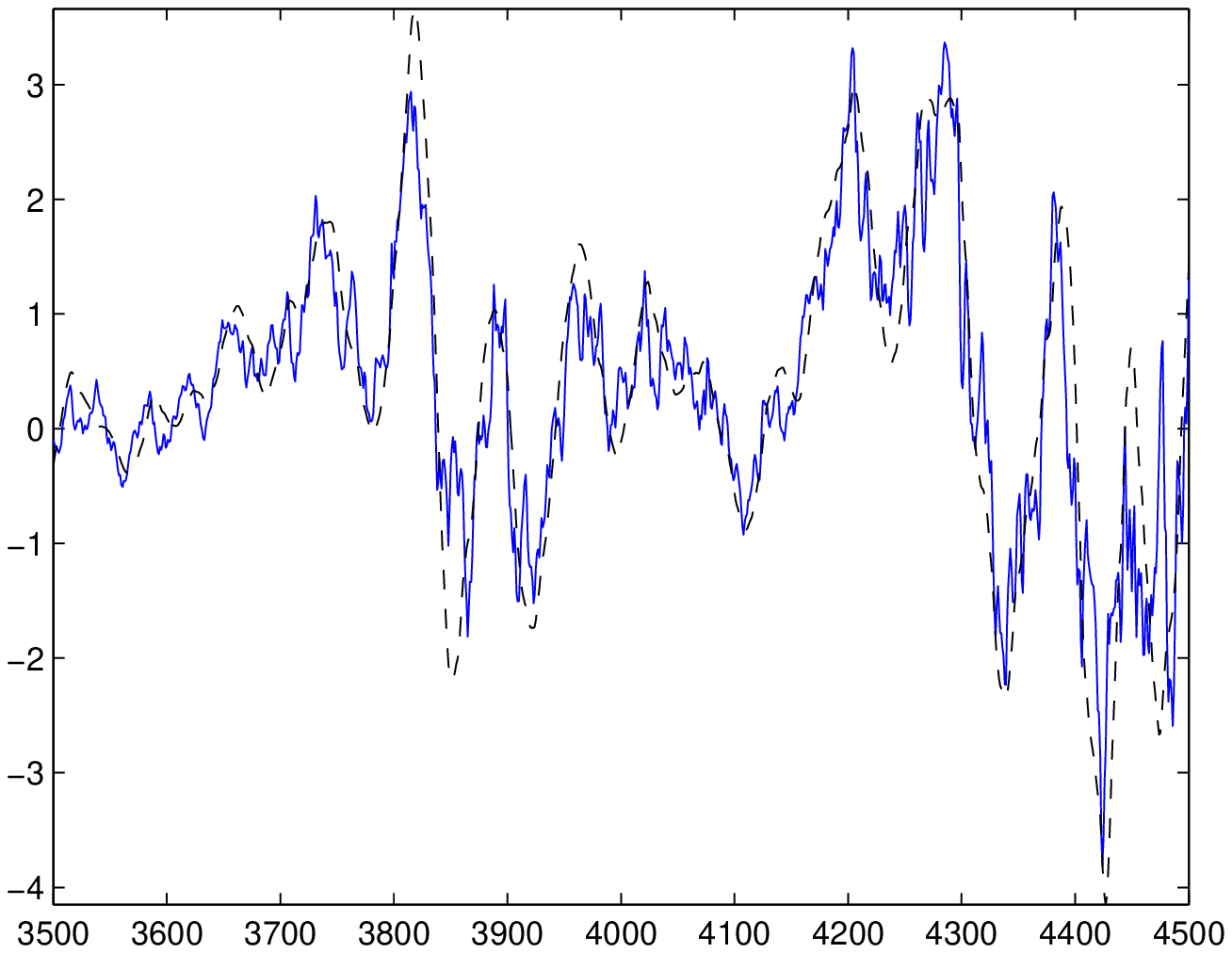}}}}}
\caption{D\'{e}rivations num\'{e}riques \label{fig_diff}}
\end{figure*}

La figure \ref{fig_diff} compare deux approches:
\begin{itemize}
\item l'une obtenue, comme souvent en analyse technique, en deux \'{e}tapes:
\begin{itemize}
\item une moyenne mobile sur 50 jours,
\item une diff\'{e}rence finie pour la d\'{e}rivation,
\end{itemize}
\item l'autre, qui demande davantage de points, obtenue
selon une m\'{e}thodologie d\'{e}duite du {\S} \ref{par_debest}.
\end{itemize}
La premi\`{e}re est bruit\'{e}e, ce qui n'est pas le cas de la seconde qui,
soulignons-le, ne s'accompagne pas de retard suppl\'{e}mentaire.
\subsubsection{Estimation des fluctuations rapides}
\begin{figure*}[!ht]
\centering {\subfigure[\footnotesize $X_{\tiny{\rm fluctuation}}(t)$
d\'{e}duit d'une moyenne glissante classique (--) et $X_{\tiny{\rm
fluctuation}}(t)$ propos\'{e} (- -) ]{
\rotatebox{-0}{\resizebox{!}{6.5cm}{%
   \includegraphics{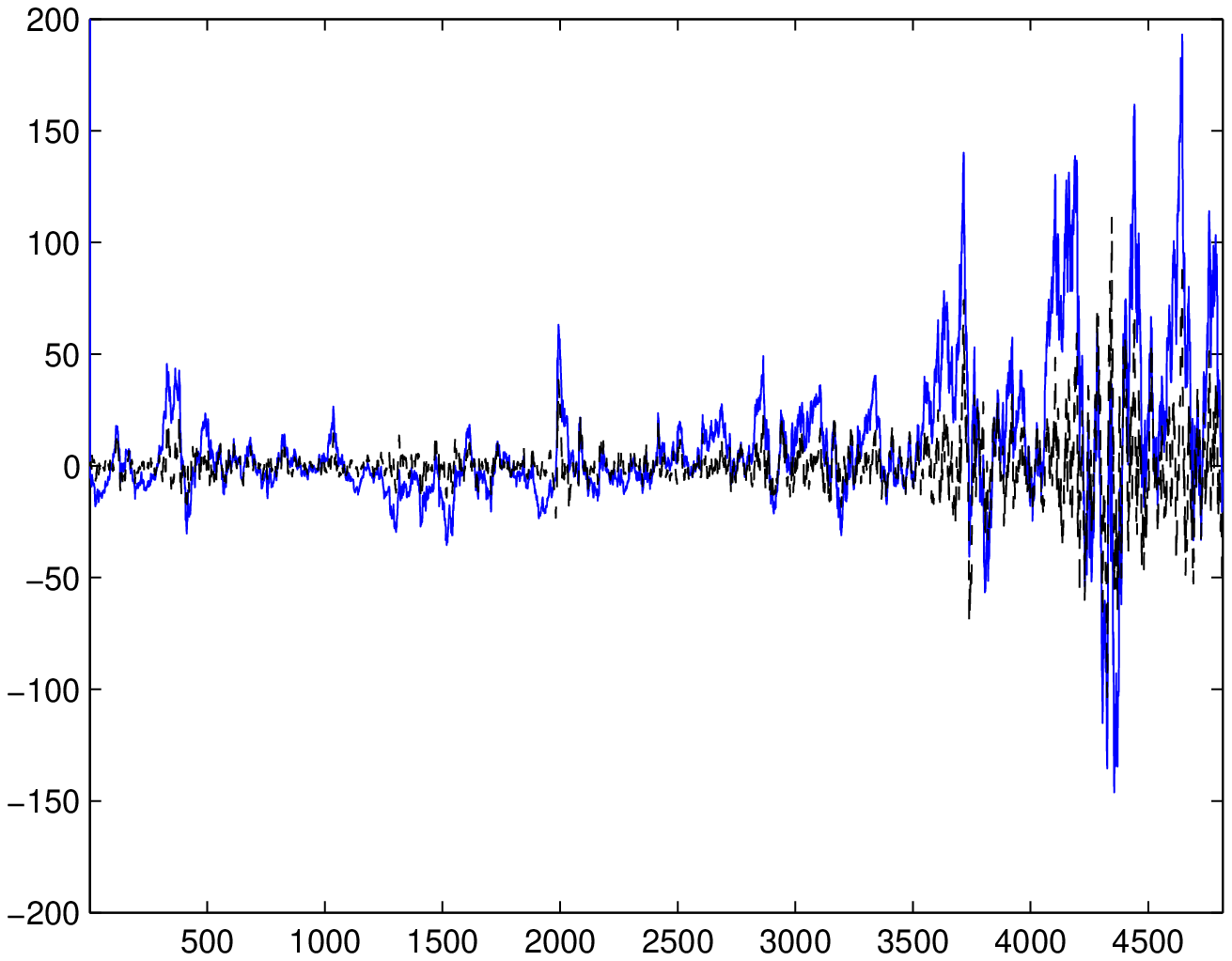}}}}}
{\subfigure[\footnotesize zoom de la figure \ref{fig_vol}-(a)]{
\rotatebox{-0}{\resizebox{!}{6.5cm}{\includegraphics{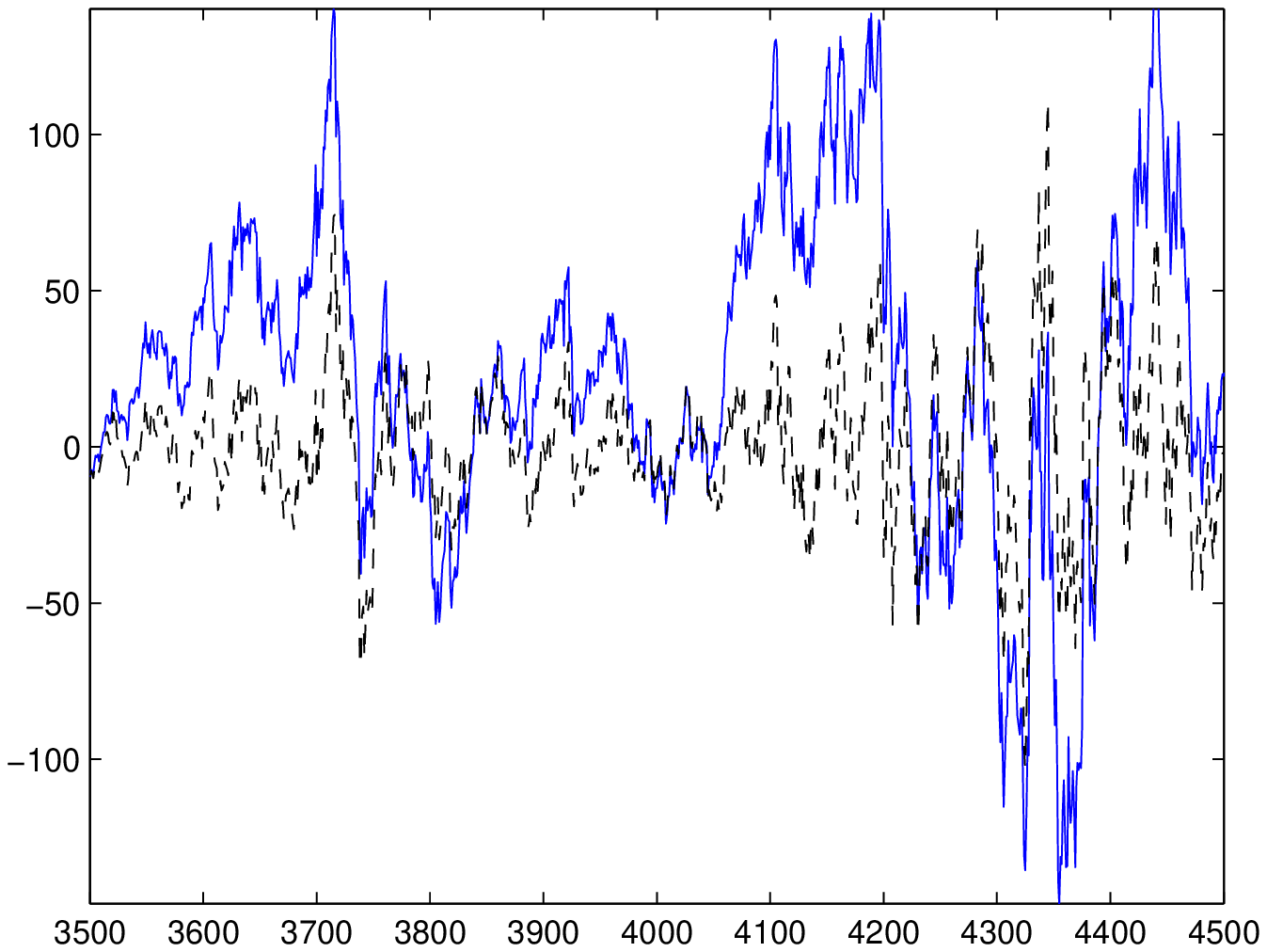}}}}}
\caption{Estimation des fluctuations rapides \label{fig_vol}}
\end{figure*}

La figure \ref{fig_vol} repr\'{e}sente deux estimations des fluctuations
rapides:
\begin{itemize}
\item L'une, calcul\'{e}e gr\^{a}ce \`{a} une moyenne glissante de $100$ jours,
classique,
fournissant l'esp\'{e}rance, ne fluctue pas vraiment autour de $0$, car
elle poss\`{e}de une composante \og basse fr\'{e}quence \fg.
\item L'autre, calcul\'{e}e par nos techniques, est bien meilleure. Elle
devrait servir \`{a} construire des indicateurs d'\emph{arr\^{e}t de
perte}\footnote{\textit{Stop loss}, en anglais.}.
\end{itemize}

\subsubsection{Pr\'{e}diction de volatilit\'{e}}
\begin{figure*}[!ht]
\centering {\subfigure[\footnotesize Rendement logarithmique instantan\'{e}
 $r_{\Delta T=1} ({\mathfrak{A}}) (t)$ ]{\rotatebox{-0}{\resizebox{!}{6.5cm}{\includegraphics{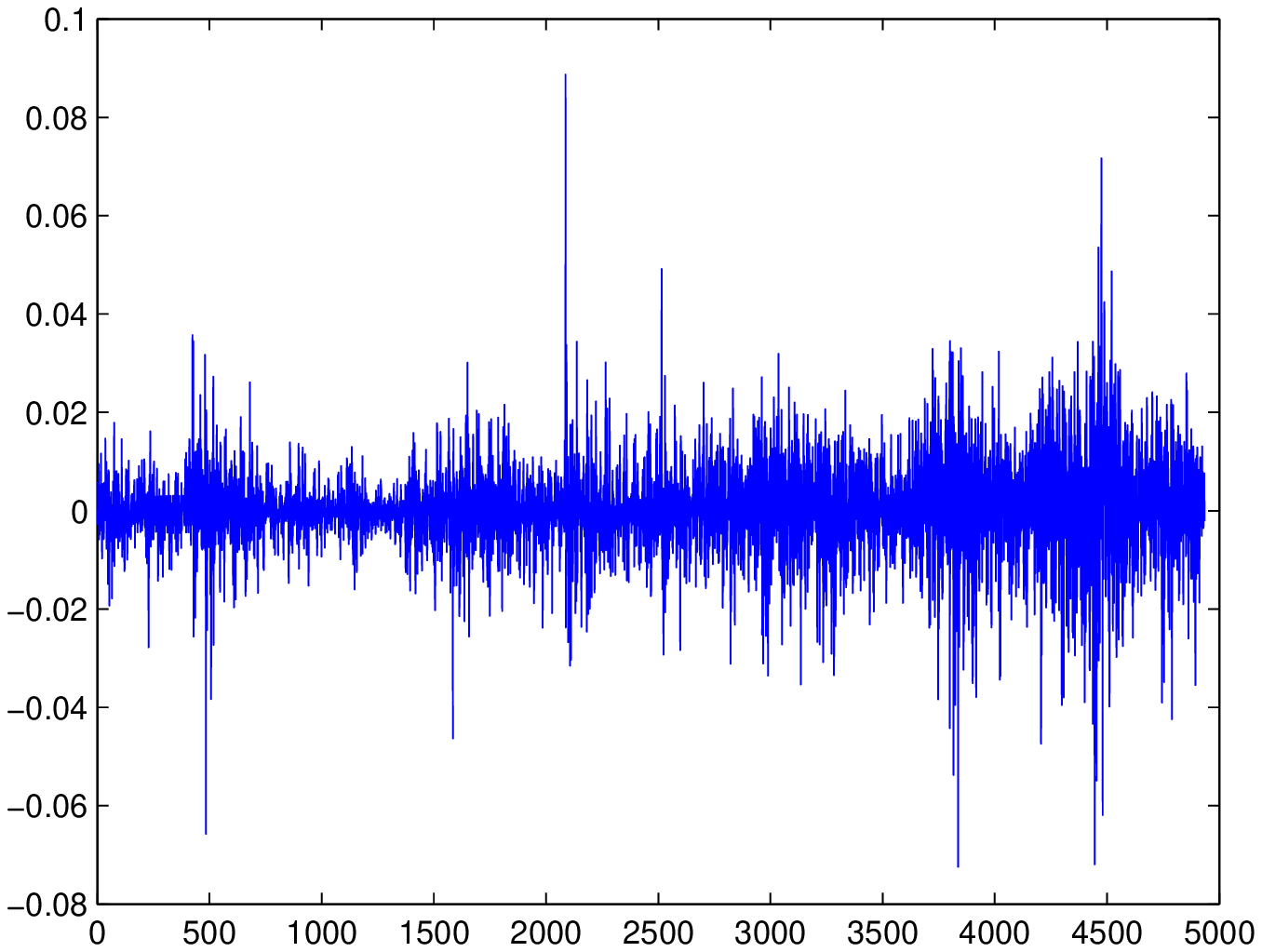}}}}}
{\subfigure[\footnotesize Rendement logarithmique normalis\'{e} $\bar{r}_{\Delta T=500} ({\mathfrak{A}}) (t)$]{
\rotatebox{-0}{\resizebox{!}{6.5cm}{\includegraphics{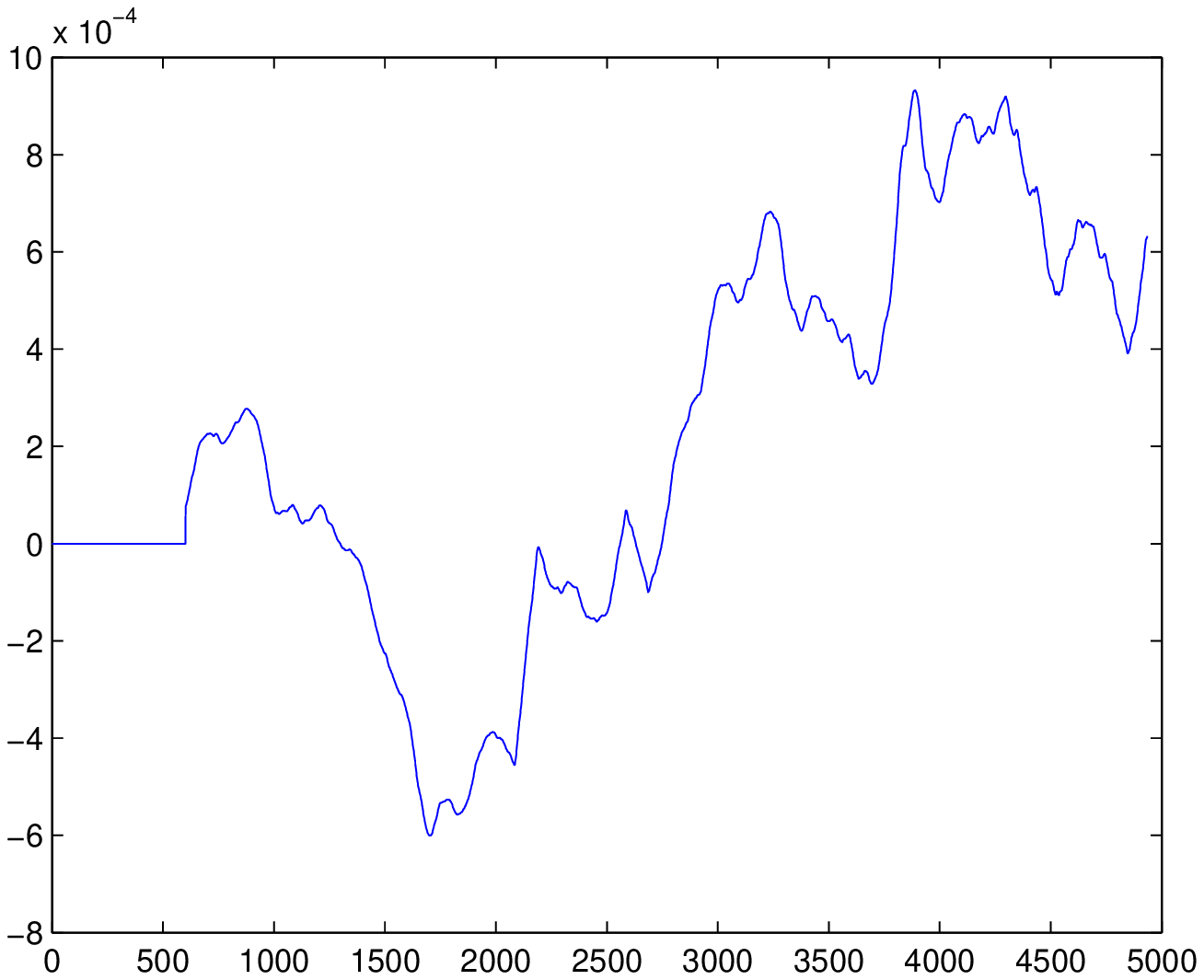}}}}}
\caption{Exemple de calculs des rendements \label{fig_rend}}
\end{figure*}

\begin{figure*}[!ht]
\centering {\subfigure[\footnotesize Volatilit\'{e}  ${\text{\bf vol}}_{\Delta T=500} ({\mathfrak{A}})(t) $ (--) et sa pr\'{e}diction \`{a} 20 jours (- -)
 ]{\rotatebox{-0}{\resizebox{!}{6.5cm}{\includegraphics{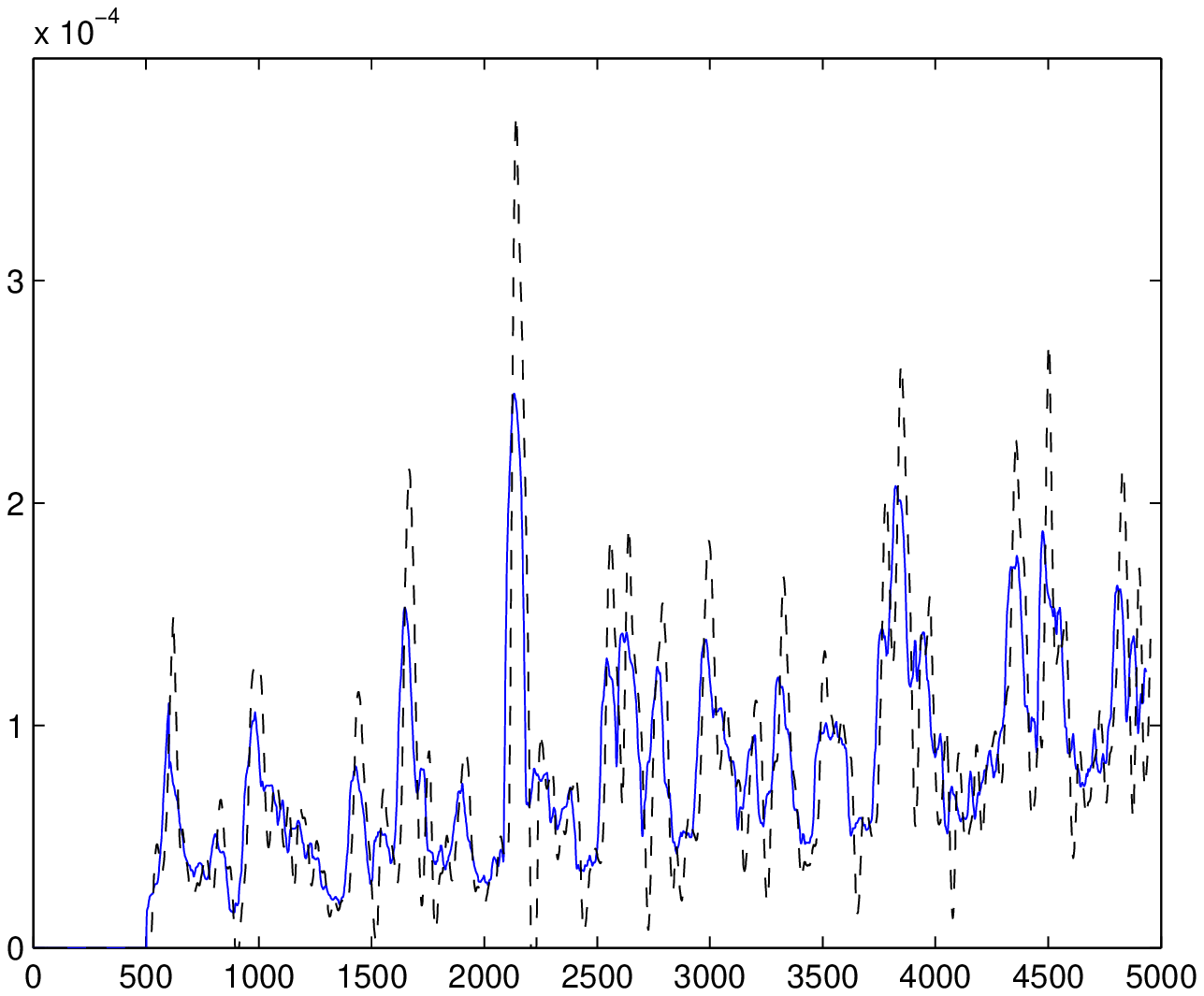}}}}}
\caption{Volatilit\'{e} du rendement normalis\'{e} \label{fig_volpred}}
\end{figure*}

On remarque une diff\'{e}rence notable entre les figures
\ref{fig_rend}-(a) et \ref{fig_rend}-(b), repr\'{e}sentant les
rendements de l'or avec deux intervalles de temps. La figure
\ref{fig_volpred} donne la volatilit\'{e} \eqref{vola} et une pr\'{e}diction
\`{a} $20$ jours, calcul\'{e}e par une m\'{e}thode standard d'interpolation. Les
r\'{e}sultats semblent plus prometteurs que ceux obtenus avec les
m\'{e}thodes de type \emph{ARCH/GARCH}, populaires depuis Engle (voir
\cite{engle}).


\begin{remarque}
La d\'{e}tection de \emph{ruptures} de \cite{rupt} a d\'{e}j\`{a} \'{e}t\'{e} employ\'{e}e
en \cite{malo2,beta,delta,douai} pour fournir des pr\'{e}visions
prometteuses de changements brutaux. Le manque de place nous emp\^{e}che
de reprendre, ici, ces calculs.
\end{remarque}

\section{Gestion dynamique de portefeuilles  et de
strat\'{e}gies}\label{dyn}
\subsection{Pr\'{e}liminaires\label{GDF_pre}}
La valeur d'un portefeuille $\mathfrak{P}$, \`{a} $N$ actifs
$\mathfrak{A}_i$, $i = 1, \dots, N$, de valeurs $P_i(t)$ \`{a} l'instant
$t \geq 0$, est
\begin{equation}\label{portf}
P(t) = \sum_{i = 1}^{N} x_i(t) P_i(t)
\end{equation}
Le choix des quantit\'{e}s $x_i(t)$, $x_i (t) \geq 0$, d'actifs
$\mathfrak{A}_i$, $i = 1, \dots, N$, assure la gestion dynamique du
portefeuille. On suppose $x_i(t)$ sans fluctuations rapides. Donc
\begin{equation}\label{E}
E(\prod_{i = 1}^{N} \left(x_i P_i)^{\nu_i}\right)(t) = \prod_{i =
1}^{N} (x_i(t))^{\nu_i} (E (P_i)^{\nu_i})(t)
\end{equation}
\subsection{Am\'{e}lioration dynamique des performances\label{GDF_am}}
On cherche \`{a} augmenter le ratio de Sharpe \eqref{sharp} du
portefeuille $\mathfrak{P}$. Supposons, pour simplifier les calculs,
que pendant l'intervalle $\Delta T$, fix\'{e}, les quantit\'{e}s $x_i(t)$
restent constantes. D'apr\`{e}s les formules \eqref{aver}, \eqref{vola},
\eqref{portf} et \eqref{E}, il est loisible de consid\'{e}rer ce ratio
comme fonction de $t$, $x_i (t)$, $i = 1, \dots, N$, fonction que
nous ne chercherons pas \`{a} \'{e}crire explicitement\footnote{Une telle
\'{e}criture deviendrait ais\'{e}e en rempla\c{c}ant le rendement logarithmique
par l'arithm\'{e}tique.}. On associe \`{a} ce ratio un syst\`{e}me de
coordonn\'{e}es $t, x_{1}, \dots, x_{N}, y$ et l'hypersurface
${\mathfrak{shr}}_{\Delta T}(\mathfrak{P})$, dite de Sharpe, d\'{e}finie
par $y = \text{shr}_{\Delta T}({\mathfrak{P}})$. Une hypoth\`{e}se,
naturelle, de d\'{e}rivabilit\'{e} locale des tendances permet de d\'{e}terminer
le plan tangent en un point courant de ${\mathfrak{shr}}_{\Delta
T}(\mathfrak{P})$. Introduisons, alors, les estim\'{e}es de
$\frac{\partial y}{\partial x_{i}}$, $i = 1, \dots, N$, calcul\'{e}es
selon les techniques de \cite{beta}.

Notons $d x_i$ l'accroissement \og infinit\'{e}simal \fg, c'est-\`{a}-dire
\og petit \fg, de $x_i$, $i = 1, \dots, N$. Faisons, pour
simplifier, l'hypoth\`{e}se d'\'{e}viter tout effet de
\emph{levier}\footnote{\emph{Leverage}, en anglais.}, qui se traduit
par
\begin{equation*}\label{levier}
P_1(t)dx_1 + \dots + P_N(t)dx_N = 0
\end{equation*}
Voici une esquisse des r\`{e}gles permettant l'am\'{e}lioration dynamique
des performances:
\begin{itemize}
\item Si $\frac{\partial y}{\partial x_{i}} > 0$ (resp. $< 0$), on
choisit $dx_i > 0$ (resp. $<0$).
\item Si $\frac{\partial y}{\partial x_{i_1}} > 0$ et
$\frac{\partial y}{\partial x_{i_2}} > 0$ et si $\frac{\partial
y}{\partial x_{i_1}} \gg \frac{\partial y}{\partial x_{i_2}}$, on
fait cro\^{\i}tre plut\^{o}t $x_{i_1}$.
\item Si $\frac{\partial y}{\partial x_{i_1}} < 0$ et $\frac{\partial
y}{\partial x_{i_2}} < 0$ et si $\frac{\partial y}{\partial x_{i_1}}
\ll \frac{\partial y}{\partial x_{i_2}}$, on fait d\'{e}cro\^{\i}tre
plut\^{o}t $x_{i_1}$.
\item Si $\frac{\partial y}{\partial x_{i}} \simeq 0$, on choisit $dx_i =
0$, \`{a} moins que les autres d\'{e}riv\'{e}es partielles soient n\'{e}gatives, ce
qui conduit \`{a} prendre $dx_i > 0$.
\end{itemize}

\begin{remarque}
Ces r\`{e}gles de bon sens, que l'on peut affiner \`{a} loisir en
\begin{itemize}
\item prenant d'autres crit\`{e}res,
\item am\'{e}liorant plusieurs crit\`{e}res simultan\'{e}s,
\end{itemize}
abandonnent volontairement la recherche d'un optimum global.
\end{remarque}

\begin{remarque}
Les calculs pr\'{e}c\'{e}dents se g\'{e}n\'{e}ralisent imm\'{e}diatement \`{a} un choix
dynamique entre plusieurs strat\'{e}gies.
\end{remarque}


\subsection{Illustrations num\'{e}riques\label{ill_num2}}
\begin{figure*}[!ht]
\centering {\subfigure[\footnotesize Evolution des poids]{
\rotatebox{-0}{\resizebox{!}{6.5cm}{%
   \includegraphics{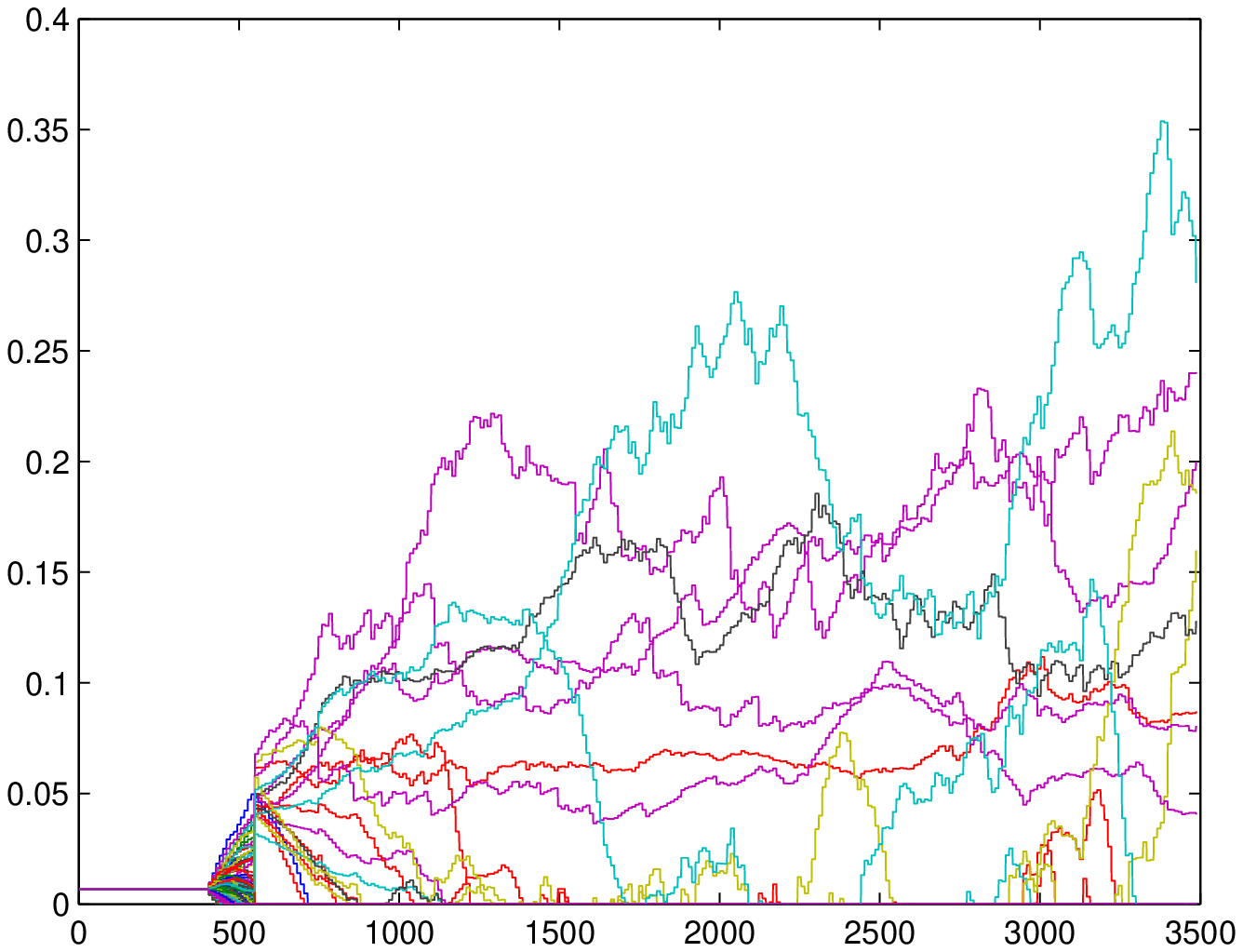}}}}}
{\subfigure[\footnotesize Valeur du portefeuille avec et sans
gestion dynamique]{
\rotatebox{-0}{\resizebox{!}{6.5cm}{\includegraphics{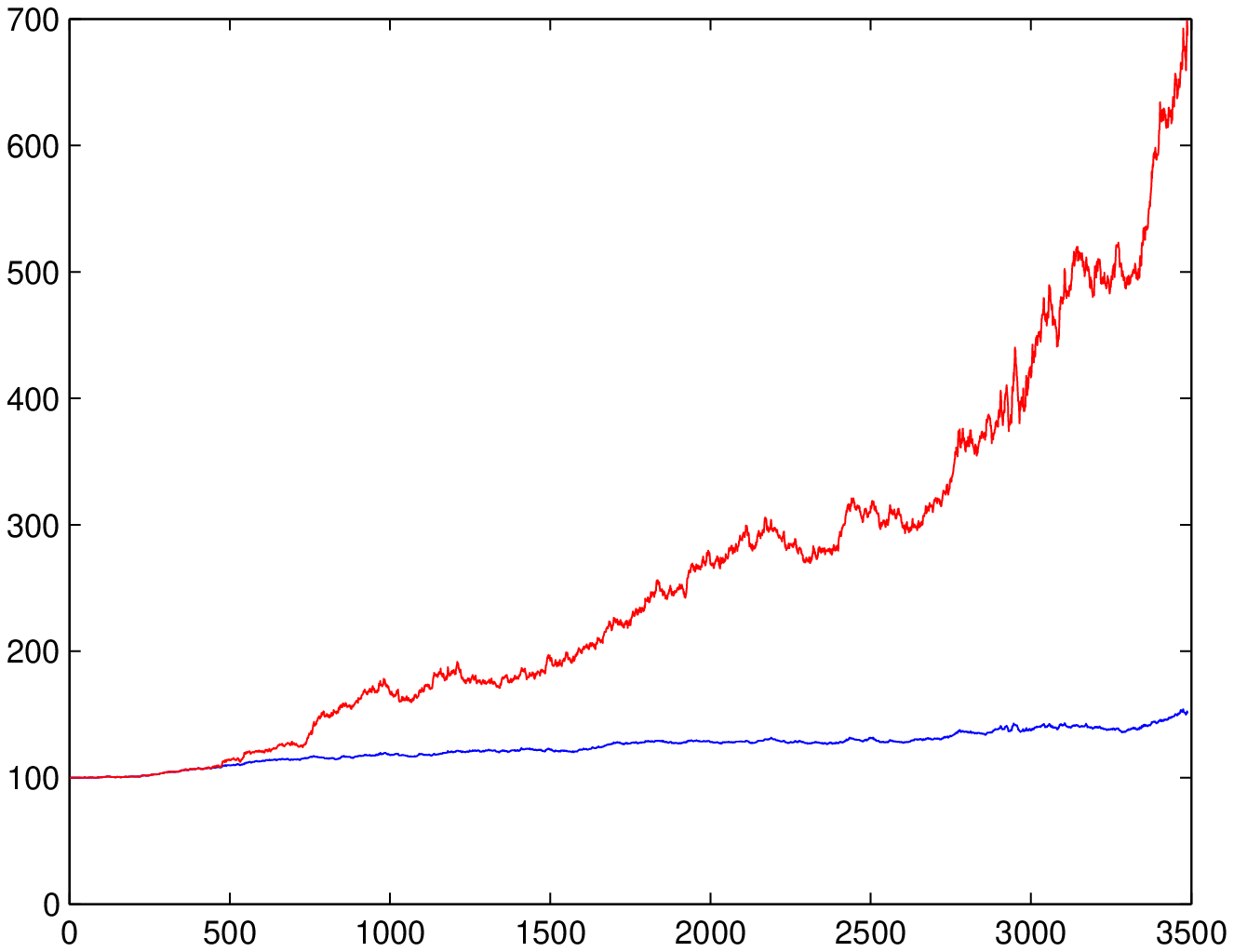}}}}}
\caption{Gestion d'un portefeuille de 38 futures \label{fig_GDF}}
\end{figure*}

On consid\`{e}re, du 28 janvier 1997 au 7 d\'{e}cembre 2010,
\begin{itemize}
\item un portefeuille compos\'{e}
de 38 futures, SPX, Dax, Hsi, Nky, Ndx, Kospi, EUR, GBP, AUD, NZD,
CHF, Cl1, NG1, HO1, Golds, Silver, LMCADS03, C1, W1, S1, RR1, TY1,
RX1, OE1, DU1, JPY, US1, FV1, TU1, LC1, SMI, UKX, G1, SPTSX, Pall,
CC1, CT1, SB1;
\item quatre strat\'{e}gies  pour chaque actif.
\end{itemize}
On remplace les poids $x_i (t)$ de \eqref{portf}  par $x_{i}^{j}
(t)$, o\`{u} $i$ d\'{e}note l'actif et $j$ la strat\'{e}gie. Ici, $1 \leq i \leq
38$, $1 \leq j \leq 4$:  l'espace de configuration est de dimension
152. Les graphiques \ref{fig_GDF}-(a) et \ref{fig_GDF}-(b)
indiquent, respectivement, l'\'{e}volution de ces nouveaux poids et
l'\'{e}volution des gains, qui sont de 15\% par an, en moyenne, avec un
ratio de Sharpe de 1.5 par an, en moyenne. Les
baisses\footnote{\textit{Drawdowns}, en anglais.}, ou pertes, ne
d\'{e}passent pas 10\%.

\begin{remarque}
Le portefeuille initial, en $t = 0$, est \'{e}quir\'{e}parti, c'est-\`{a}-dire
\textit{a priori} non optimal. Plut\^{o}t qu'utiliser les techniques
actuelles d'optimisation statique, lourdes comme d\'{e}j\`{a} dit en
introduction, il semble pr\'{e}f\'{e}rable de recourir \`{a} la d\'{e}marche
pr\'{e}c\'{e}dente. Les donn\'{e}es historiques permettent de d\'{e}marrer nos
calculs en $t < 0$ avec un portefeuille \'{e}quir\'{e}parti. Le r\'{e}sultat
obtenu en $t = 0$ est le portefeuille initial.
\end{remarque}

\section{Conclusion}\label{conclusion}
C'est la majeure part de la finance quantitative que nous comptons
embrasser \`{a} terme. L'absence de mod\`{e}les probabilistes \textit{a
priori} devrait mener \`{a} des m\'{e}thodes plus simples et efficaces.

Les mod\`{e}les probabilistes ne tiennent pas seulement une place, tr\`{e}s
exag\'{e}r\'{e}e \`{a} notre avis, en \'{e}conom\'{e}trie et en finance, mais aussi en
automatique et en signal, domaines plus traditionnels de
l'ing\'{e}nierie et des math\'{e}matiques appliqu\'{e}es. Nous nous contenterons
dans cette courte conclusion d'\'{e}voquer les filtres de Kalman
(\cite{kal1,kal2}) car ils jouent aussi un r\^{o}le en \'{e}conom\'{e}trie
(voir, par exemple, \cite{gourieroux,hamilton}) et, donc, en finance
(voir, par exemple, \cite{wells}). En effet, ils
\begin{itemize}
\item sont tributaires d'une mod\'{e}lisation pr\'{e}cise non seulement de la
dynamique, mais aussi de la statistique des bruits;
\item exigent, comme les correcteurs PID en automatique
industrielle, un r\'{e}glage d\'{e}licat des gains\footnote{Un avantage
formidable des PID \emph{intelligents}, issus de la commande sans
mod\`{e}le (\cite{modelfree1,marseille}), est un r\'{e}glage facile.}.
\end{itemize}
C'est pourquoi les \emph{reconstructeurs d'\'{e}tat} de \cite{rec}, o\`{u}
peu importe la statistique des bruits, devraient, les remplacer
avantageusement\footnote{Cette affirmation s'applique aussi aux
\emph{observateurs asymptotiques}, familiers en automatique.}, si
l'on a foi en un mod\`{e}le de la dynamique! Ajoutons que ces
reconstructeurs se g\'{e}n\'{e}ralisent sans difficult\'{e}, en utilisant les
outils du {\S} \ref{par_debest}, au non-lin\'{e}aire \cite{nl}. Ce n'est
pas le cas, on ne le sait que trop, des filtres de Kalman.

\begin{remarque}
Le filtre de Kalman est employ\'{e}, par exemple, pour estimer le fameux
coefficient b\^{e}ta (voir, par exemple, \cite{bodie,roncalli}), fourni
par un mod\`{e}le lin\'{e}aire tr\`{e}s contest\'{e}. Plut\^{o}t qu'utiliser nos
reconstructeurs, mieux vaut sans doute adopter l'approche sans
mod\`{e}le de \cite{beta,douai}.
\end{remarque}

\end{document}